\documentclass{article}

\usepackage{ijcai18}
\usepackage{times}
\usepackage{amssymb}
\usepackage{amsmath}
\usepackage{balance}
\usepackage[utf8]{inputenc}
\usepackage{complexity}
\usepackage[small]{caption}
\usepackage{tikz}

\newtheorem{mytheorem}{Theorem}

\newtheorem{myobservation}{Observation}

\newcommand{\myproof}{\noindent {\bf Proof.\ \ }}

\newcommand{\myqed}{\mbox{$\diamond$}}
\newcommand{\myOmit}[1]{}

\newcommand{\mygini}{\textsc{Gini}}
\newcommand{\myefgini}{\textsc{subjective Gini}}
\newcommand{\myef}{\textsc{Envy}}

\begin{document}

\title{Fair Division Minimizing Inequality}
\author{Martin Aleksandrov \\ TU Berlin \\ martin.aleksandrov@tu-berlin.de \And Cunjing Ge \\ Chinese Academy of Sciences \\ gecj@ios.ac.cn \And Toby Walsh \\ 
Data61, CSIRO  \\
toby.walsh@data61.csiro.au}


\maketitle

\begin{abstract}
Behavioural economists have shown that people are often averse to inequality and will make choices to avoid unequal outcomes. In this paper, we consider how to allocate indivisible goods fairly so as to minimize inequality. We consider how this interacts with axiomatic properties such as envy-freeness, Pareto efficiency and strategy-proofness. We also consider the computational complexity of computing allocations minimizing inequality. Unfortunately, this is computationally intractable in general so we consider several tractable greedy online mechanisms that minimize inequality. Finally, we run experiments to explore the performance of these methods. 
\end{abstract}

\section{Introduction}\label{sec:intro}

In resource allocation, one of
the most frequently used normative measures of fairness is envy-freeness
(no agent envies another's allocation). Unfortunately, when the
resources are indivisible, envy-free allocations may \emph{not} 
exist. In addition, computing an envy-free allocation when it exists
is computationally intractable. 
Another desirable property in resource allocation is Pareto efficiency. In contrast 
to envy-free allocations, Pareto efficient allocations \emph{always} exists and can
be computed quickly. However, Pareto efficient allocations may not be 
very fair (e.g.\ giving all items to a single agent is Pareto efficient).
We consider here whether minimizing the inequality between agents offers an alternative to envy-freeness and Pareto efficiency for the fair division of indivisible items. 
A number of different measures of inequality have been proposed in economics 
(e.g. \cite{gini1912,atkinson1970,hoover1936}). We focus on the Gini
index as it has been commonly used in many other settings. However, 
it would be interesting to consider other measures such as the Atkinson,
and Hoover (aka Robin Hood) indices. 

\emph{Our results:} We start our paper with a motivating example. We consider three normative inequality measures for fair division: the \emph{Gini} index, the \emph{subjective Gini} index and the \emph{envy} index. These three indices measure the quality of allocations and mechanisms between perfect equitability and envy-freeness. Unlike envy-free allocations which may \emph{not} exist, allocations that minimize these three measures \emph{always} exist. We study the relationship between the Gini, subjective Gini and envy indices and envy-freeness, Pareto efficiency and strategy-proofness. For example, we show that there are fair division problems when \emph{none} of the envy-free allocations minimizes the inequality indices. We further study the complexity of computing allocations minimizing each of these indices. Unfortunately, most of these computational problems are intractable. For this reason, we propose three tractable online mechanisms that allocate each item in a given sequence thus minimizing the three inequality indices without the knowledge of the future items in the sequence. We finally run experiments with these online mechanisms.

\section{Formal background}\label{sec:form}

We consider a fair division problem with $n$ agents and $m$ indivisible
items. Each agent has some private cardinal \emph{utility} $u_{i}(o_j) \in\mathbb{Q}^{\geq 0}$ for each item $o_j$ but can submit a public cardinal \emph{bid} $v_{i}(o_j)\in\mathbb{Q}^{\geq 0}$ for each item $o_j$. An \emph{instance} of a fair division problem thus has (1) \emph{agents}
$a_1,\ldots,a_n$, (2) indivisible \emph{items} $o_1,\ldots,o_m$ and (3) a bid matrix $(v_{i}(o_j))_{n\times m}$. Let $A$ be an allocation of items to agents. We write $A_i$ for the bundle of items allocated to agent $i$, and $u_i(B)$ for the utility to agent $a_i$ of the items in the bundle $B$. We assume additive utilities. That is, $u_{i}(B)=\sum_{o_j \in B} u_{i}(o_j)$. In economics, incomes and wealth are additive for the population. Also, in a food bank, donated products are additive for the bank. Additivity offers an elegant compromise between simplicity and expressivity in our model as well as in many other theoretical models (e.g.\ \cite{bevia1998,brams2003,chevaleyre2008,keijzer2009,lesca2010}). 

We consider welfare, fairness and efficiency notions. The \emph{utilitarian welfare} of $A$ is equal to $\sum_{i\in[1,n]} u_i(A_i)$. The \emph{egalitarian welfare} of $A$ is equal to $\min_{i\in[1,n]} u_i(A_i)$.
An allocation $A$ is \emph{envy-free} iff $u_i(A_i)\geq u_{i}(A_j)$ for every $i, j$. An allocation $A$ is \emph{Pareto efficient} iff there is no
other allocation $A^{\prime}$ such that $\forall i : u_i(A_i^{\prime})\geq u_i(A_i)$ and $\exists k : u_k(A_k^{\prime})>u_k(A_k)$. We further consider only \emph{responsive} mechanisms that compute an allocation of items to agents based on their positive bids. A desirable property of mechanisms is that they cannot be manipulated. 
A mechanism is \emph{strategy-proof} if, for each instance, an agent
cannot increase their utility by misreporting their bids. We are interested in properties of the actual ex post outcomes returned by mechanisms.

One of the most frequently used measures of inequality is the \emph{Gini} index. It is
commonly used to measure inequality in income or wealth.
The Gini index satisfies a number of desirable properties such as anonymity, scale independence, population independence,
and the transfer principle (inequality reduces when we take from the rich and 
give to the poor). We will use it here
to measure inequality between agents in the utility of the items allocated to them. 
More precisely, the Gini index of an allocation equals half of the relative mean absolute difference in utilities of the agents. 
\begin{eqnarray*}
 \mbox{Gini} & = & 
\frac{\sum_{i=1}^n \sum_{j=1}^n |u_i(A_i)-u_j(A_j)|}{2\sum_{i=1}^n\sum_{j=1}^n u_i(A_i)}  \\
& = & \frac{\sum_{i=1}^n \sum_{j=1}^n |u_i(A_i)-u_j(A_j)|}{2n\sum_{i=1}^n u_i(A_i)} 
\end{eqnarray*}
The Gini index lies in the interval [0,1], taking the value 0 when 
all $n$ agents get the same utility, and $1-\frac{1}{n}$ when all but one agent get
zero utility. In a plot of the cumulative distribution, the Gini index measures 
the ratio of the area that lies between the line of equality (i.e.\ all $n$ agents get the same utility) and the Lorenz curve \cite{endriss2013}.

\section{A motivating example}\label{sec:mot}

A simple example provides some motivation.
Suppose Alice, Bob and Carol arrive at the
car hire office and are offered to rent 
a Renault, a Skoda, or a Toyota car. 
Alice knows that Skoda's share 
their mechanicals with VW, and likes
reliable German cars, so she prefers
the Skoda most. Bob is torn between the
Skoda and the more unusual Renault. And Carole loves quirky cars,
so has a strong preference for the Renault. She is also
an environmentalist, so dislikes VW and has a strong preference against
the Skoda. 
Their precise utilities for the different
cars are given in the following table. 
Who gets what car?

\begin{table}[h]
\begin{center}
\begin{tabular}{|c|c|c|c|} \hline
& Renault & Skoda & Toyota \\ \hline
Alice & 1 & 8 & 3 \\
Bob & 8 & 7 & 1 \\
Carol & 18 & 1 & 8 \\ \hline
\end{tabular}
\end{center}
\end{table}

There is no envy-free allocation. Bob and Carol
both most prefer the Renault and only one of
them can get it. The allocation with 
the least amount of envy (either of one person for
another or in total) allocates the Renault to Carol, the Skoda
to Bob and the Toyota to Alice. 
This is also the optimal allocation from 
a welfare perspective with both the maximum
utilitarian and egalitarian welfare. 
However, Alice
might not consider this allocation
fair as she gets less than half the utility of Bob or Carol,
as well as less than half the utility of her most preferred 
car, whilst Carol gets her most preferred car and Bob gets
a car with value close to his maximum utility. 

We might decide instead that it is fairer to chose
from amongst those 
allocations which minimize the inequality
between Alice, Bob and Carol. 
For instance, allocating the Renault to Bob, the Skoda to Alice
and the Toyota to Carole is one such allocation. 
Everyone gives their car the same 8 units of utility.
This allocation is Pareto efficient and has a Gini index of zero, the minimum possible. In this allocation, only Carol envies Bob, but 
since she gets as much utility for her car
as both Alice and Bob get for their cars, this might be 
acceptable. 

Note that there is another allocation that minimizes
inequality. Allocating the Renault to Alice, the Skoda
to Carol and the Toyota to Bob gives everyone 
the same 1 unit of utility. This also has 
a Gini index of zero. 
However, everyone now has their least
preferred car, and everyone envies everyone else. Moreover, this allocation is not Pareto efficient and has the minimal 
welfare possible, both from the utilitarian and egalitarian perspective. 

To sum up, this example suggests
that whilst the Gini index can help in choosing
between allocations, we cannot minimize inequality alone. Amongst
allocations that minimize inequality, we 
might look to maximize welfare, minimize
envy, etc. Minimizing inequality does, however, have an advantage over
envy-freeness as a primary measure of fairness. An allocation of
indivisible items minimizing inequality always exists whilst an envy-free allocation may not. 

\section{The subjective Gini index}\label{sec:subgini}

As remarked earlier, the Gini index is typically used to measure inequality in income and wealth
distribution. However, we are concerned here with the distribution of indivisible items not money, and
importantly agents can have different \emph{subjective} utilities for these items. 
For example, the utility you get for an item is not necessarily the same as the utility
I get for it. 

Should it increase the ``inequality'' of an allocation
that someone else gets an item they value when you 
have little or even no value for it? 
To return to our motivating example, suppose Alice gets the Renault, 
Bob gets the Toyota, and Carol gets the Skoda.
Everyone gets 1 unit of utility so this allocation
has a Gini index of zero. 
But from everyone's subjective perspective, this is not 
a very equitable allocation of items. 
For instance, from Alice's perspective, 
rather than the 1 unit of utility she gets,
she would get 8 units of utility for
Carol's car and 3 for Bob's. 
And from Bob's perspective, 
rather than the 1 unit of utility he gets,
he would get 8 units of utility for
Alice's car and 7 for Carol's. 

We propose the \emph{subjective Gini} index to take such
differences into consideration. 
We modify the definition of the Gini index to sum the difference
in utility an agent has for its allocation and the utility the \emph{same}
agent has for the allocation of items to other agents. 
\begin{equation*}
\mbox{subjective Gini} =  \frac{\sum_{i=1}^n \sum_{j=1}^n |u_i(A_i)-u_{i}(A_j)|}{2 \sum_{i=1}^n\sum_{j=1}^n u_{i}(A_j)} 
\end{equation*}
Like the Gini index, the subjective Gini index is between [0,1] taking the value 0 when 
each agent gives the same utility to each bundle of items, and $1-\frac{1}{n}$ when one
agent gets all items. Returning again to our motivating example, the 
allocation in which each agent gets 1 unit of utility 
has a Gini index of 0 but a subjective Gini index
of 23/55 (=0.41818181818). 
The allocation in which each agent gets 8 units of utility
might be more preferred as it has a lower subjective
Gini index of
37/110 (=0.33636363636). 

\section{The envy index}\label{sec:efindex}

Minimizing the subjective Gini index will find
allocations which divide the items into bundles
so that each bundle has similar utility for each agent. 
This reminds us of a fairness concept
such as the maximin share when each agent’s utility should be at least as high
as the agent can guarantee by dividing the items into as many bundles as there are players and receiving
their least desirable bundle \cite{bjpe2011}.

On the plus side, an allocation
which minimizes the subjective Gini index always exists, unlike maximin fair shares \cite{pwec14}. 
On the negative side, such an allocation may not be envy-free. To overcome this, 
we propose also an \emph{envy index}
whose definition is closely related to that of the subjective Gini index. This
new index is focused on the amount of envy in an allocation. 
Minimizing this index will return an envy-free allocation when it exists. 

\begin{equation*}
\mbox{envy} = \frac{\sum_{i=1}^n \sum_{j=1}^n \max\lbrace 0,u_{i}(A_j)-u_{i}(A_i)\rbrace}{\sum_{i=1}^n \sum_{j=1}^n u_{i}(A_j)} 
\end{equation*}

The envy index is between [0,1] taking the value 0 when 
the allocation is envy-free, and
tending towards 1 as we increase the number of agents and allocate all items to just one agent. 
It is easy to see that the envy index is never greater (and sometimes smaller) than
the subjective Gini index. Returning to our motivating example, the unique allocation minimizing the envy with index of 6/110 (=0.05454545454) 
allocates the Renault to Carol, the Skoda
to Bob and the Toyota to Alice. 
As we noted, this is also the optimal allocation from 
a welfare perspective with both the maximum
utilitarian and egalitarian welfare. 

\section{Relationship to envy-freeness}\label{sec:ef}

We consider how these indices relate to a fairness concept such as envy-freeness. 
Suppose that an envy-free allocation exists. Clearly, such an allocation minimizes the envy index.
On the other hand, envy-free allocations may not minimize the Gini or 
subjective Gini indices. 

\begin{mytheorem}\label{thm:one}
There exist problems with envy-free allocations
on which no envy-free allocation minimizes the Gini or
subjective Gini index. 
\end{mytheorem}

\myproof
Consider 2 agents and 2 items. 
Suppose the first agent gives the first item
a utility of 1 and
the second a utility of 2,
whilst the second agent
gives utilities of 3 and 1 respectively.
The only envy free 
allocation gives the first item to the second agent and
the second item to the first agent. However, the unique allocation
that minimizes the Gini index gives the first item to the
first agent and the second item to the 
second agent. In this allocation, 
both agents envy each other.

Consider 3 agents and 3 items. 
Suppose the first agent has 
a utility of 9, 1 and 5 for the items respectively,
the second agent has a utility of 5, 9 and 1
respectively, and the third agent has a utility
of 1, 5, and 9 respectively. Then the
unique envy-free allocation gives each agent their
most valued item. 
However, the unique allocation that minimizes
the subjective Gini index gives each agent
their second most preferred item, i.e.\ the one they value with utility of 5. 
\myqed

The examples in the proof of Theorem~\ref{thm:one} critically depend on
the agents not sharing common utilities for items. 
When utilities are common, there is no
incompatibility between envy-freeness and
minimizing the Gini or subjective Gini indices. 
If an allocation is envy-free and agents have common
utilities, then every agent must get the same utility
for every bundle of items.

\begin{myobservation}\label{obs:one}
With common utilities, an allocation is envy-free iff the Gini and subjective Gini indices are
zero.
\end{myobservation}

\section{Relationship to Pareto efficiency}\label{sec:pe}

Another fundamental notion in fair division
is Pareto efficiency. We would prefer allocations
where no agent can improve their outcome without
making others worse off. 
Pareto efficiency is not necessarily compatible
with minimizing inequality. 
The first example in the proof of Theorem~\ref{thm:one}
shows that Pareto efficiency and the Gini index are incompatible. This should perhaps not be
surprising as other fairness properties
are also incompatible with Pareto efficiency. For example, an allocation that is envy free
may not necessarily be Pareto efficient. Moreover, each envy-free allocation can be Pareto dominated only by allocations that are not envy-free \cite{keijzer2009}. 
It follows quickly that minimizing the envy index is not compatible with 
Pareto efficiency. We can show that the same is true for the subjective Gini index. 
 
\begin{mytheorem}\label{thm:two}
There exist problems on which no Pareto efficient allocation minimizes the subjective Gini index. 
\end{mytheorem}

\myproof
Consider 2 agents and 4 items. 
Suppose the first agent gives items
$o_1,o_3$ a utility of 1, item $o_2$ a utility of $2-\epsilon$ and
$o_4$ a utility of $\epsilon$,
whilst the second agent
gives utilities of $2-\epsilon,1,\epsilon,1$ to $o_1,o_2,o_3,o_4$ respectively.
Then the only allocation minimizing the subjective Gini index 
allocates $o_1,o_3$ to the first agent, and
$o_2,o_4$ to the second agent. However, the only Pareto efficient allocation swaps items $o_1,o_2$, giving $o_1$ to the
second agent, and $o_2$ to the 
first agent. 
\myqed

Again, with common utilities, there is no
incompatibility between Pareto efficiency and
minimizing the Gini, subjective Gini and envy indices. This follows because each allocation, including those that minimize these indices, is Pareto efficient.

\begin{myobservation}\label{obs:two}
With common utilities, any allocation minimizing the Gini, subjective Gini or envy index is Pareto efficient. 
\end{myobservation}

We can measure the trade-off between Pareto efficiency and minimizing 
one of these indices. The \emph{egalitarian/utilitarian price of an index} for a given welfare is the
ratio between 
the best welfare of any Pareto efficient 
allocation 
and the worst welfare of an
allocation minimizing the index.

\begin{mytheorem}\label{thm:three}
The utilitarian and egalitarian prices of the Gini and subjective Gini
indices are unbounded. 
\end{mytheorem}

\myproof
Consider 2 agents, 2 items and let $\epsilon < \frac{1}{2}$. 
Suppose the first agent gives item
$o_1$ a utility of $\epsilon$ and
$o_2$ a utility of $1-\epsilon$,
whilst the second agent
gives utilities of $2-\epsilon$ and $\epsilon$ respectively.
Then the Pareto efficient outcome with the best utilitarian and
egalitarian welfare 
allocates $o_1$ to the second agent, and
$o_2$ to the first agent. However, the only allocation
that minimizes the Gini index does the reverse. 
The egalitarian price of the Gini index is then 
$\frac{1-\epsilon}{\epsilon}$ which is unbounded
as $\epsilon$ goes to zero.
The utilitarian price is 
$\frac{3-2\epsilon}{2\epsilon}$ which is unbounded
as $\epsilon$ goes to zero.
The same example demonstrates
that the utilitarian and egalitarian price
of the subjective Gini index are also unbounded.
\myqed

For the envy index, we have examples 
where the utilitarian price grows as the
number $n$ of agents. We conjecture that
this may also be an upper bound. For 
the egalitarian price, we can show that the
price is unbounded. 

\begin{mytheorem}\label{thm:four}
The egalitarian price of the envy
index is unbounded. 
\end{mytheorem}

\myproof
Consider 3 agents, and 3 items. 
Suppose the first agent gives a utility of 1 to each item, 
and both the second and third agents gives 
utilities of 8, 4, and 4 respectively to the 3 items. 
The Pareto efficient outcome with
the best egalitarian welfare
allocates the item with utility 8 to
the second or third agent, 
and each of the remaining items to
one of the other agents. This has
an egalitarian welfare of 1 unit. 
However, the allocation that minimizes
the envy index gives 
the item with utility 8 to
the second agent, both the 
other items to the third agent,
or vice versa. As the first agent gets no items,
this has an egalitarian welfare
of zero units. Hence, the egalitarian price of
the envy index is unbounded. 
\myqed

\section{Relationship to strategy proofness}\label{sec:sp}

If we use a mechanism that minimizes one of these indices,
agents have an incentive to declare false utilities. 
Again, this should not be too surprising. We often need to
choose between fairness and strategy-proofness.
For example, the random priority is
strategy-proof but it can return allocations 
which are not envy-free \cite{bogomolnaia2001}.

\begin{mytheorem}\label{thm:five}
A mechanism
which minimizes the Gini, subjective Gini or envy index
is not strategy proof.
\end{mytheorem}

\myproof
For the Gini index, consider the first example from proof of Theorem~\ref{thm:one}. 
If agents sincerely report their utilities,
the first agent gets $o_1$ and the second
agent gets $o_2$. 
If the first agent misreports their
utilities as $1/2$ and 3 respectively, 
the agents swap items, and both
agents are better off. 
Similarly if the second agent misreports
their utilities as 2 and $1/2$ respectively,
the agents swap items, and both
agents are better off. 

For the subjective Gini index, consider 2
agents and 4 items. Let the first agent have utilities
$u_{11}=1,u_{12}=3/2,u_{13}=1,u_{14}=1/2$ whereas the second agent have
utilities $u_{21}=3/2,u_{22}=1,u_{23}=1/2,u_{24}=1$. Suppose sincere
play. The mechanism that minimizes the subjective Gini index gives to
each agent both items for which they have utility 1, or both items for
which they have utility $3/2$ and $1/2$. The expected utility of each
agent is then $2$. Suppose next that the first agent reports utilities
1, $3/2$, 0, 0 respectively. The mechanism now gives the first and
second items to the first agent and the third and fourth items to the
second item. The utility of the first agent increases to $5/2$. 

For the envy index, we can use the same instance as for the subjective Gini index.
\myqed

\section{Computational complexity}\label{sec:com}

In this section, we turn our attention to computational properties of the
Gini, subjective Gini and envy indices. Computing envy-free
allocations is $\NP$-hard even with just 2 agents, and common
utilities \cite{bouveret2008}. It immediately follows that finding an
allocation minimizing
the envy index is $\NP$-hard. 
The proof from \cite{schneckenburger2017} 
showing that minimizing the Atkinson index is $\NP$-hard can be reused to prove
that finding an allocation that minimizes the Gini or subjective Gini
index is NP-hard. 

One way to deal with this intractability is to use algorithms that are fast enough for small values of $n$ or $m$ \cite{bliem2016}. Another way is to identify some tractable cases. For example, with $n$ agents and $n$ items, minimizing the subjective Gini or envy index is polynomial. Each envy-free allocation (whenever it exists) minimizes the envy index. Each envy-free allocation with common utilities (whenever it exists) minimizes the subjective Gini index. Interestingly, minimizing the Gini index is also polynomial in this case. For each utility value $u$, consider the instance in which only the utilities equal to $u$ are left. Each envy-free allocation in this instance minimizes the Gini index. Computing allocations minimizing the indices in this setting with $n$ agents and $n$ items takes $O(n^{5/2})$ time \cite{hopcroft1973}. 

\section{Online mechanisms}\label{sec:on}

Another approach to deal with the intractability
of computing allocations that minimize inequality or envy
is to use greedy online mechanisms. These will often return an allocation with little inequality or envy, even if there is no guarantee that it is minimal. Online mechanisms are also applicable when
the allocation problem is itself online \cite{aleksandrov2017eom,mehta2013,walsh2017}. We consider three online randomized mechanisms. These mechanisms can be applied to an offline problem by picking an (perhaps random) order of the items. WLOG, let $o=(o_1,\ldots,o_m)$ be such an order. Each mechanism computes a set of agents feasible for each next $o_j$ in $o$ given an allocation $A_{j-1}$ of $o_1$ to $o_{j-1}$. A feasible agent then receives $o_j$ with probability that is uniform with respect to the other feasible agents.

\begin{itemize}
\item \mygini: this decides that $a_i$ is feasible for $o_j$ if $v_i(o_j)>0$ and $A_{j-1}\cup\lbrace (a_i,o_j)\rbrace$ minimizes the Gini index

\item \myefgini: this decides that $a_i$ is feasible for $o_j$ if $v_i(o_j)>0$ and $A_{j-1}\cup\lbrace (a_i,o_j)\rbrace$ minimizes the subjective Gini index

\item \myef: this decides that $a_i$ is feasible for $o_j$ if $v_i(o_j)>0$ and $A_{j-1}\cup\lbrace (a_i,o_j)\rbrace$ minimizes the envy index
\end{itemize}

A powerful technique to study online mechanisms is competitive
analysis \cite{sleator1985}. This has recently been applied to online 
fair division \cite{aleksandrov2017mcm}. Competitive analysis identifies the loss in efficiency due to the data arriving in an online fashion. An online mechanism $M$ is \emph{$c$-competitive} for a given welfare $w$ iff there exists a constant $b$ such that, whatever the order $o$ of items, $w(\mbox{OPT})\leq c\cdot w(M,o)+b$ holds where $w(M,o)$ is the welfare of $M$ on $o$ and $w(\mbox{OPT})$ is the optimal offline welfare. 

A mechanism that is $c$-competitive has a ratio $c$. Most of the ratios of our mechanisms are unbounded. For example, we can use the instance from the proof of Theorem 10 in \cite{aleksandrov2015ijcai} and show that both the utilitarian and egalitarian ratios of \myefgini\ are unbounded. We next prove similar results for \mygini\ and \myef.

\begin{mytheorem}\label{thm:six}
The utilitarian and egalitarian competitive ratios of \mygini\ are unbounded. 
\end{mytheorem}

\myproof
For \mygini, consider the online fair division of items $o_1,o_2$ to agents $a_1,a_2$. Let the first agent have a utility 1 for $o_1$ and $\epsilon$ for $o_2$ whilst the second agent have a utility $\epsilon$ for $o_1$ and 1 for $o_2$ where $\epsilon>0$. The mechanism allocates $o_1$ to $a_2$ and $o_2$ to $a_1$ and thus returns utilitarian and egalitarian welfares of $2\epsilon$ and $\epsilon$. The optimal offline allocation allocates $o_2$ to $a_2$ and $o_1$ to $a_1$ and thus returns utilitarian and egalitarian welfares of $2$ and $1$. The competitive ratios are equal to $\frac{1}{\epsilon}$ which goes to $\infty$ as $\epsilon$ goes to zero. 
\myqed

\begin{mytheorem}\label{thm:seven}
The utilitarian competitive ratio of \myef\ is at least $\frac{n}{2}$ whilst its egalitarian competitive ratio is unbounded. 
\end{mytheorem}

\myproof
For the utilitarian ratio, consider $n$ agents and $n$ items. Let the first agent 
have utility $n$ for each item, and each other agent have utility 1 for
each item. Then \myef\ will allocate the first item to the first
agent, and then each subsequent item to a new agent. 
The utilitarian welfare of this allocation is $2n-1$. The 
optimal utilitarian welfare is $n^2$. 

For the egalitarian ratio, consider the online fair division of items $o_1,o_2$ to agents $a_1,a_2$. Let the first agent have a utility 1 for each item whilst the second agent have a utility $\epsilon$ for $o_1$ and 0 for $o_2$ where $\epsilon>0$. The mechanism allocates both items to the first agent, and thus returns an egalitarian welfare of $0$. The optimal offline allocation gives to each agent an item they like, and returns egalitarian welfare of $\epsilon$. The egalitarian ratio is $\infty$.
\myqed

We can also measure the price of anarchy of these online mechanisms. The \emph{price of anarchy} is closely related to the competitive ratio but now supposing agents act strategically \cite{koutsoupias1999,aleksandrov2015ijcai}. The \emph{price of anarchy} of an online mechanism for a given welfare is the ratio between the best welfare of an allocation when agents are sincere and the worst welfare of an allocation when agents are strategic. Interestingly, the price of anarchy of each of our online mechanisms is at least to $n$. We conjecture that this may also be their upper bound.

\begin{mytheorem}\label{thm:eight}
The utilitarian and egalitarian prices of anarchy of \mygini, \myefgini\ and \myef\ are at least $n$.
\end{mytheorem}

\myproof
Consider an instance with $n$ agents and $n$ items. For $i\in\lbrace 1,\ldots,n\rbrace$, let $a_i$ has utility of $1$ for $o_i$, and utility of $\epsilon>0$ for each other item. The optimal offline allocation gives to each $a_i$ their most valued item. The utilitarian and egalitarian welfares of this allocation are $n$ and $1$ respectively. 

We start with \mygini. At round 1, this mechanism gives the first item to one of the agents who likes with it $\epsilon$. The first agent then has an incentive to report $\epsilon$ for this item simply because they do not know what items will arrive next. By a similar argument, at round 2, the optimal play for the second agent is to bid $\epsilon$, and so on for each other round. At the end of the allocation, each agent gets expected utility of $\frac{1}{n}+\frac{(n-1)}{n}\epsilon$. The utilitarian and egalitarian welfares of this strategic allocation go to $1$ and $\frac{1}{n}$ respectively as $\epsilon$ goes to zero. The prices are consequently at least $n$. 

We next consider \myefgini. The sincere play is optimal for each agent with this mechanism because they get each item with probability $\frac{1}{n}$. The welfares go to $1$ and $\frac{1}{n}$ respectively as $\epsilon$ goes to zero. The prices are at least $n$. 

We finally consider \myef. This mechanism tends to allocate each item to agents with the highest utility for this item. By similar arguments as for \mygini, we conclude that the optimal play of each agent is to bid 1 for each item. Each agent thus gets expected utility of $\frac{1}{n}+\frac{(n-1)}{n}\epsilon$.
\myqed

Despite the fact that these mechanisms are not competitive supposing agents act sincerely, they become more competitive supposing agents act strategically. Moreover, each of
these mechanisms does as well as any other online mechanism at minimizing their
respective
index. An online mechanism $M_1$ is \emph{ex post optimal} for a given index iff, for each other online mechanism $M_2$, each online problem and each allocation $A_2$ returned by $M_2$, there exist an allocation $A_1$ returned by $M_1$ such that the index of $A_2$ is at least the same as the index of $A_1$. We show ex post optimality only for \mygini. The proof for the other two mechanisms can similarly be done by analogy.

\begin{mytheorem}\label{thm:nine}
The \mygini\ mechanism is ex post optimal for the Gini index. 
\end{mytheorem}

\myproof
Suppose that \mygini\ is not optimal. Hence, there is another online mechanism $M$, an online problem and an allocation $A_M$ such that the index of $A_M$ is strictly lower than the minimum index of an allocation $A_{\mygini}$ returned by \mygini. This means that there is a round $j\in(1,m]$ at which $A_M$ and $A_{\mygini}$ differ for item $o_j$ but coincide for items $o_1$ to $o_{j-1}$. Let $A_{j-1}$ denote the allocation of $o_1$ to $o_{j-1}$ in $A_M$ and $A_{\mygini}$. WLOG, let $M$ allocate $o_j$ to $a_1$ whereas \mygini\ allocate it to $a_2$ given $A_{j-1}$. We have that the Gini index of $A_{j-1}\cup\lbrace (o_j,a_1)\rbrace$ with $M$ is lower than the Gini index of this allocation with \mygini. Hence, \mygini\ does not minimize this index given $A_{j-1}$. This is a contradiction with the definition of \mygini.
\myqed

Finally, Theorems~\ref{thm:six} and~\ref{thm:seven} suggest that, in the worst-case, these
online mechanisms have performance that cannot be bounded, whereas 
the Theorem~\ref{thm:eight} suggests that no other online mechanism can do better.

\section{Experiments}\label{sec:exp}

We ran an experiment to see how these online mechanisms would perform in practice. 
We generated 100 instances of $n=5$ agents, $m\in\lbrace 10,20,30,40,50,60,70,80,90,100\rbrace$ items and integer utilities drawn uniformly at random from $\lbrace 0,1,\ldots,m\rbrace$. For each combination of $n$ and $m$, we computed the Gini index, the subjective Gini index, the envy index, the egalitarian welfare and the utilitarian welfare of 100 000 sampled allocations returned by \mygini, \myefgini\ and \myef. We report in our graphs only the average results because their standard deviations were less than 1\% of them. We further omit our results for the subjective Gini index for reasons of space. 

\begin{figure}[h]
\centering
\includegraphics[width=0.5\columnwidth,height=3.5cm,clip=true,trim=0 0 44 25]{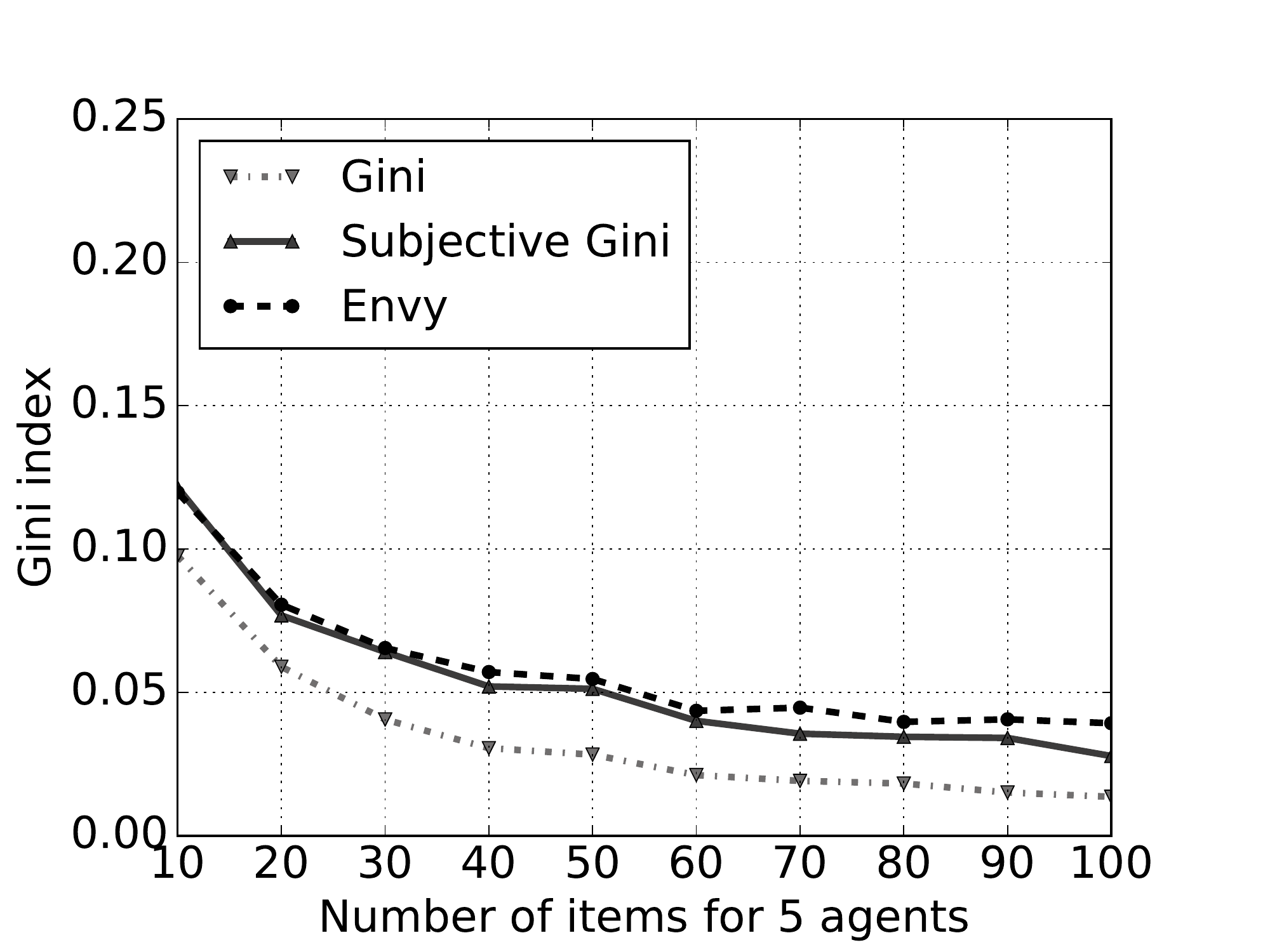}%
\includegraphics[width=0.5\columnwidth,height=3.5cm,clip=true,trim=0 0 44 25]{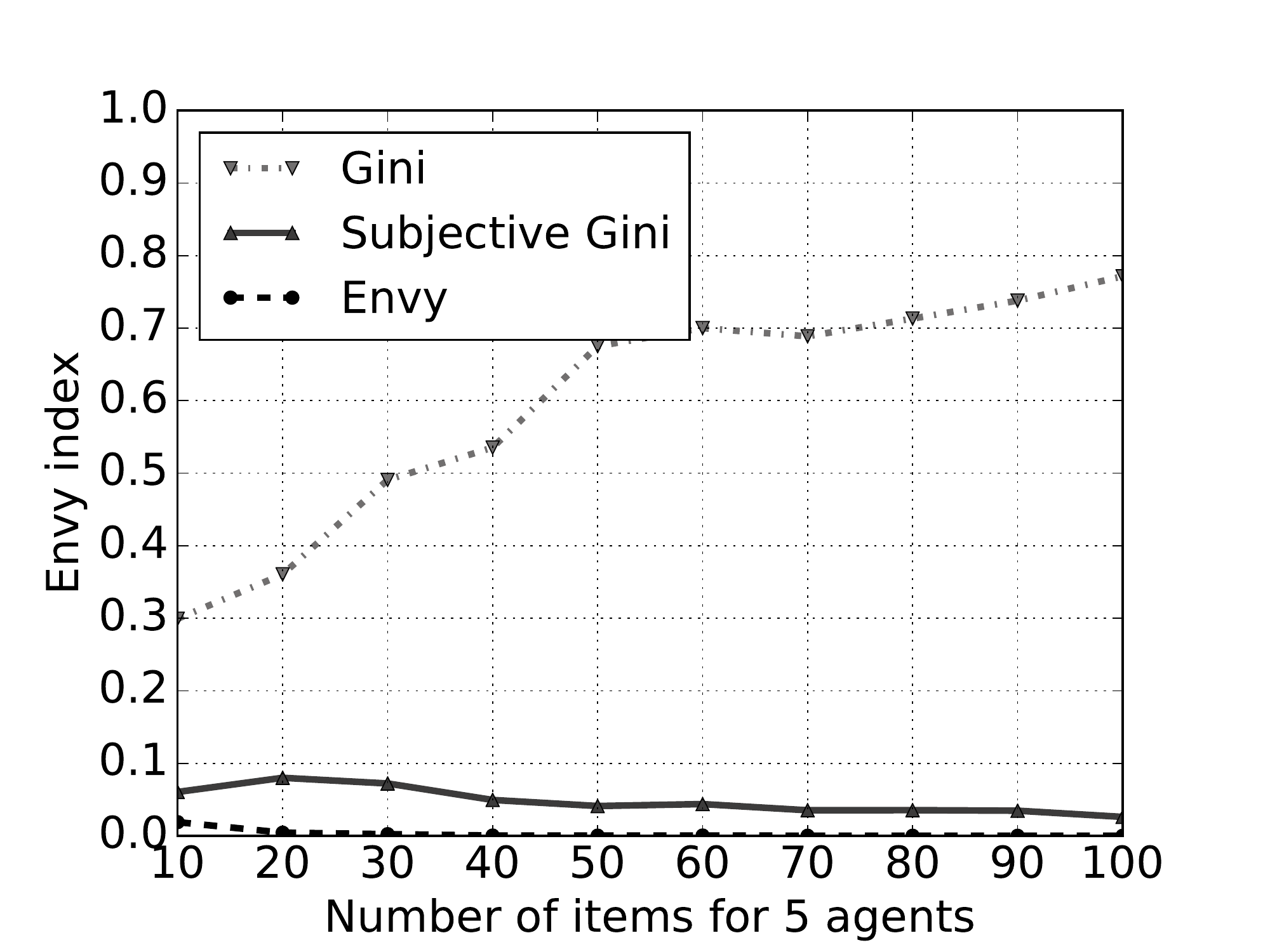}
\end{figure}

In the first graph, \mygini\ achieves the lowest value of the Gini index for each number of items. For example, the Gini value of \mygini\ is nearly 50\% lower than the Gini values of \myefgini\ and \myef\ for 100 items. This gap actually remains almost the same for any number of items in our experiment. Unfortunately, \mygini\ fails to minimize envy. In the second graph, we could clearly see that \myef\ outperforms \mygini. In fact, \myef\ achieves an envy index of almost 0 for 100 items. Interestingly, \myefgini\ tends to favor envy-freeness to equitability. Moreover, the performance of \mygini\ diverges from envy-freeness and converges to perfect equitability with more items. Perhaps, we observe this as \mygini\ tends to allocate items to agents with low utilities. In contrast, \myefgini\ and \myef\ tend to allocate items to agents with great utilities. They thus tend to minimize simultaneously both the envy and inequality.

We next report our results for the utilitarian and egalitarian ratios. The utilitarian/egalitarian ratio is the ratio between the utilitarian/egalitarian welfare returned by an online mechanism and the optimal offline utilitarian/egalitarian welfare.

\begin{figure}[h]
\centering
\includegraphics[width=0.5\columnwidth,height=3.5cm,clip=true,trim=0 0 44 25]{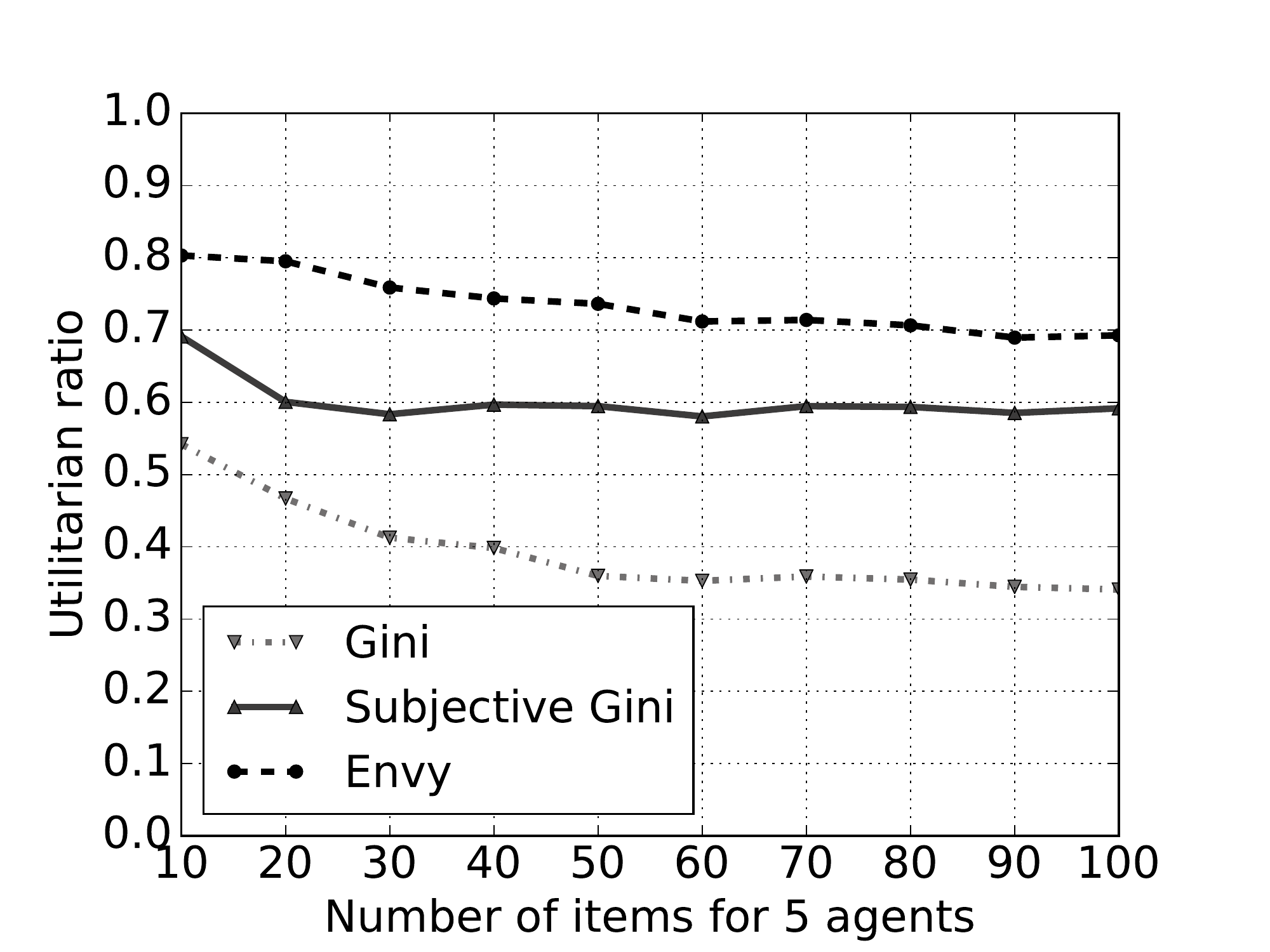}%
\includegraphics[width=0.5\columnwidth,height=3.5cm,clip=true,trim=0 0 44 25]{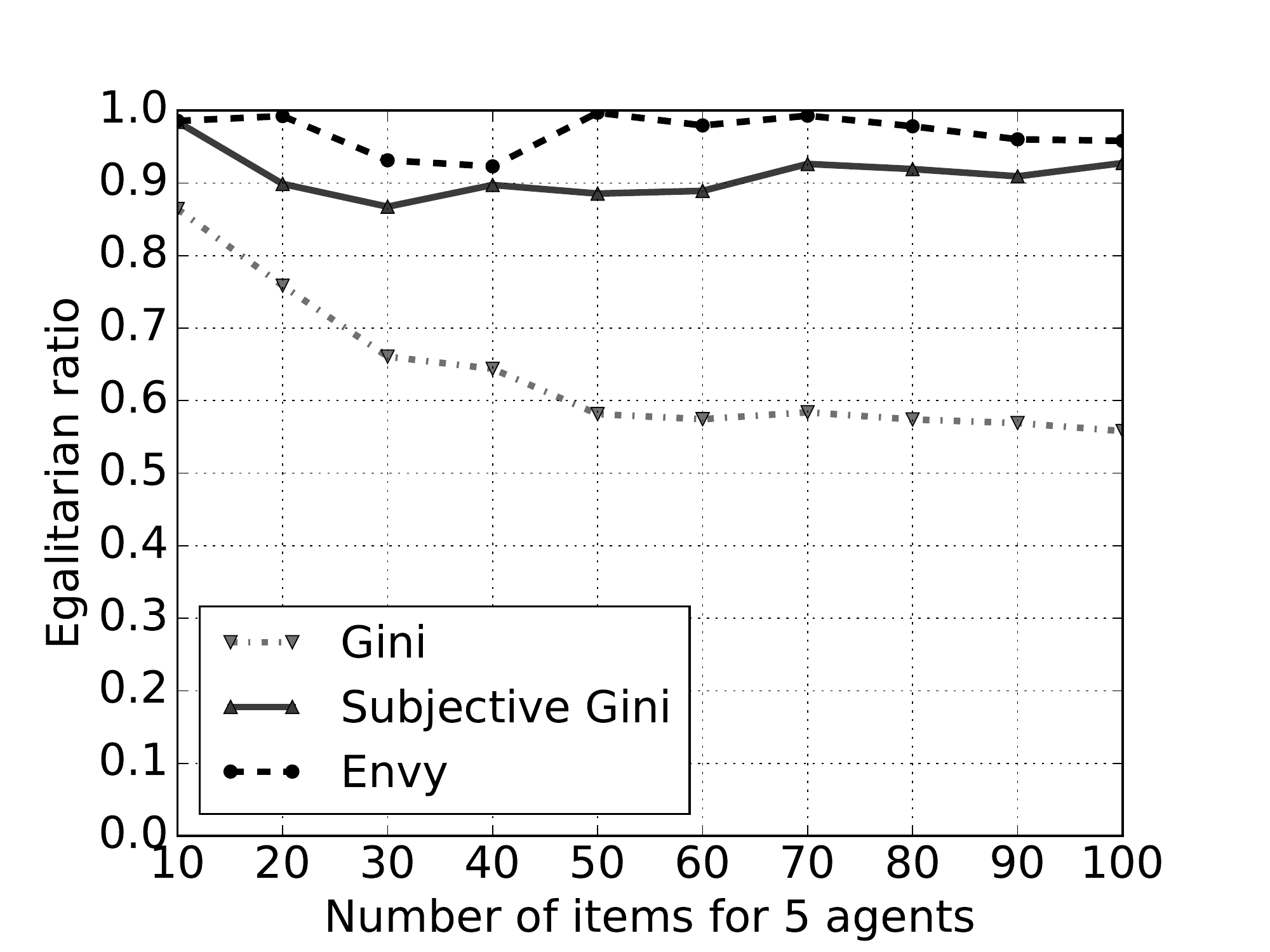}
\end{figure}

From a utilitarian perspective (the first graph), \myef\ outperforms the other two mechanisms for each number of items. For example, this mechanism achieves a utilitarian ratio close to 0.7 for 100 items. This value is nearly 16\% higher than the ratio of \myefgini\ and 100\% higher than the ratio of \mygini\ for 100 items. From an egalitarian perspective (the second graph), again \myef\ outperforms \myefgini\ and \mygini, followed closely by \myefgini. Interestingly, for each number of items, \myef\ not only minimizes the envy but also maximizes the egalitarian welfare. For 100 items, its egalitarian ratio is nearly 0.95. This value is nearly 82\% higher than the value of \mygini\ for 100 items. For both welfares, the performance of \myefgini\ is close to the performance of \myef.

Finally, our experimental results indicate that envy-freeness, equitability and welfare efficiency may be achievable in practice.

\section{Related work}\label{sec:rel}

Endriss has
formulated the task of reducing
inequality as a combinatorial optimisation problem
\cite{endriss2013}.
In particular, he studied the problem of
deciding if there exists an inequality reducing improvement such as
a Pigou-Dalton or Lorenz transfer. 
The complexity of such decision problems depends on the language used to represent
the (possibly non-additive) utilities. He also provided a modular mixed integer programming formulation that returns an allocation to minimize inequality
measures such as the Gini and Hoover indices when utilities are specified with the
XOR-language. Schneckenburger, Dorn and Endriss \cite{schneckenburger2017} consider allocating indivisible goods to minimize inequality
as measured by the 
Atkinson index. They demonstrated
that a sequence of local deals would converge on a globally optimal allocation
with the minimum Atkinson index possible, but that the number of agents and
items involved in such deals could not be bounded. For the Gini index, they
conjectured that such convergence would be very challenging if not impossible
to achieve. 

By comparison, we show that computing allocations with small inequalities might be fast in practice. Moreover, none of these works relates to other axiomatic properties. For example, \cite{aziz2015faaaa} studied 
a taxonomy of fairness concepts
related to envy-freeness and proportionality. 
However, there are fair division
problems in which even the weakest of these
concepts may not exist, whereas allocations minimizing our indices always exist. Moreover, the Gini index is characterized in \cite{perez2012}. The subjective Gini and envy indices are inspired by two measures of envy that are analysed in \cite{bosmans2018}. However, the idea of measuring envy was first proposed in \cite{feldman1974}.

\section{Conclusions}\label{sec:con}

We defined three new indices that measure the quality of allocations: the Gini, subjective Gini and envy indices. The first two indices measure
inequality within an allocation, whilst the third index measures the amount of envy. Each index could be used as a second order criterion in choosing between allocations. For example, we could choose the Pareto efficient allocation with the least value of an index. Unlike envy-free allocations which may \emph{not} exist, allocations that minimize these three indices \emph{always} exist.
We studied the relationship of these indices
with envy-freeness, Pareto efficiency and strategy-proofness. We further studied the complexity of computing allocations
minimizing each of these indices. Unfortunately, most of these
computational problems are intractable. For this reason, we proposed
three tractable online mechanisms that greedily minimize these three indices. Experiments showed that, even for modest sized problems, we 
may be able to efficiently compute allocations with limited inequality or envy as well asl with reasonably high values of the egalitarian and utilitarian welfares.

\newpage



\balance
\bibliographystyle{named}
\bibliography{inequalities}

\end{document}